\documentclass{an}
\usepackage{times}
\usepackage{graphicx}
\graphicspath{{./fig/}{./png/}}
\usepackage{fancyhdr}
\usepackage{bm}
\newcommand{\EQ}{\begin{equation}}
\newcommand{\EN}{\end{equation}}
\newcommand{\EQA}{\begin{eqnarray}}
\newcommand{\ENA}{\end{eqnarray}}
\newcommand{\eq}[1]{(\ref{#1})}

\newcommand{\Eq}[1]{Eq.~(\ref{#1})}
\newcommand{\Eqs}[2]{Eqs~(\ref{#1}) and~(\ref{#2})}

\newcommand{\Fig}[1]{Fig.~\ref{#1}}

\newcommand{\Tab}[1]{Table~\ref{#1}}

\newcommand{\bra}[1]{\langle #1\rangle}

\newcommand{\meanF}{\overline{\cal F}} 
\newcommand{\meanEMF}{\overline{\vec{\cal E}}}
\newcommand{\meanemf}{\overline{\cal E}}
\newcommand{\meanFF}{\overline{\mbox{\boldmath ${\cal F}$}} {}}

\newcommand{\meanB}{\overline{B}}

\newcommand{\meanU}{\overline{U}}

\newcommand{\meanBB}{\overline{\vec{B}}}
\newcommand{\meanJJ}{\overline{\vec{J}}}
\newcommand{\meanUU}{\overline{\vec{U}}}
\newcommand{\meanWW}{\overline{\vec{W}}}

\newcommand{\nnn}{\hat{\mbox{\boldmath $n$}} {}}

\newcommand{\zz}{\hat{\mbox{\boldmath $z$}} {}}

\newcommand{\ddelta}{{\vec{\delta}}}

\newcommand{\uu}{{\vec{u}}}
\newcommand{\BB}{{\vec{B}}}
\newcommand{\JJ}{{\vec{J}}}
\newcommand{\jj}{{\vec{j}}}

\newcommand{\bb}{{\vec{b}}}

\newcommand{\KK}{{\vec{K}}}
\newcommand{\RR}{{\vec{R}}}

\newcommand{\kk}{\mbox{\boldmath $k$} {}}

\newcommand{\nab}{\mbox{\boldmath $\nabla$} {}}
\newcommand{\OO}{\mbox{\boldmath $\Omega$} {}}
\newcommand{\oo}{\mbox{\boldmath $\omega$} {}}

\newcommand{\SSSS}{\mbox{\boldmath ${\sf S}$} {}}

\newcommand{\ii}{{\rm i}}

\newcommand{\dd}{{\rm d} {}}

\def\half{{\textstyle{1\over2}}}

\def\onethird{{\textstyle{1\over3}}}

\newcommand{\yapj}[3]{: #1, {ApJ} {#2}, #3}

\newcommand{\yan}[3]{: #1, {AN} {#2}, #3}

\newcommand{\yana}[3]{: #1, {A\&A} {#2}, #3}

\newcommand{\ypp}[3]{: #1, {PhPl} {#2}, #3}

\newcommand{\yprl}[3]{: #1, {PhRvL} {#2}, #3}
\newcommand{\ypre}[3]{: #1, {PhRvE} {#2}, #3}

\newcommand{\ymn}[3]{: #1, {MNRAS} {#2}, #3}

\newcommand{\spre}[1]{: #1, PhRvE (submitted)}

\newcommand{\yjour}[4]{: #1, {#2}, {#3}, #4}

\title{Strong mean field dynamos require supercritical helicity fluxes}
\author{Axel Brandenburg$^1$ and Kandaswamy Subramanian$^2$}
\institute{
$^1$ NORDITA, Blegdamsvej 17, DK-2100 Copenhagen \O, Denmark\\
$^2$ IUCAA, Post Bag 4, Pune University Campus, Ganeshkhind, Pune 411 007, India
}

\date{Received 2 May 2005; accepted 31 May 2005; published online 1 July 2005}
\begin{document}

\abstract{
Several one and two dimensional mean field models are analyzed
where the effects of current helicity fluxes and boundaries are included
within the framework of the dynamical quenching model.
In contrast to the case with periodic
boundary conditions, the final saturation energy of the mean field decreases
inversely proportional to the magnetic Reynolds number.
If a nondimensional scaling factor in the current helicity flux exceeds a
certain critical value, the dynamo can operate even without kinetic
helicity, i.e.\ it is based only on shear and current helicity fluxes,
as first suggested by Vishniac \& Cho (2001, ApJ 550, 752).
Only above this threshold is the current helicity flux also able
to alleviate catastrophic quenching.
The fact that certain turbulence simulations have now shown apparently
non-resistively limited mean field saturation amplitudes
may be suggestive of the current helicity flux having exceeded this
critical value.
Even below this critical value the field still reaches appreciable
strength at the end of the kinematic phase, which is in qualitative
agreement with dynamos in periodic domains.
However, for large magnetic Reynolds numbers the field undergoes subsequent
variations on a resistive time scale when, for long periods,
the field can be extremely weak.
}

\maketitle

\section{Introduction}

Astrophysically relevant dynamos tend to have boundaries or are at least
confined.
In all practically relevant cases they are certainly not homogeneous.
Exceptions are dynamos on the computer where triply-periodic
boundary conditions are used.
Despite such dynamos being so unrealistic, they have played an enormously
important role in revealing the nature of catastrophic $\alpha$ quenching
(Brandenburg 2001, hereafter referred to as B01,
Blackman \& Brandenburg 2002, hereafter referred to as BB02).
This applies in particular to the case of helically forced turbulence.
For reviews regarding recent developments see Brandenburg et al.\ (2002)
and Brandenburg \& Subramanian (2005a).
However, some important conclusions drawn from triply periodic simulations
do not carry over to the case with boundaries.
In this paper we discuss in particular the saturation field strength
and focus on the mean-field description taking the evolution equation of
current helicity with the corresponding current helicity fluxes into account.

Homogeneous turbulent dynamos saturate in such a way that the total
current helicity vanishes, i.e.\ $\bra{\JJ\cdot\BB}=0$, where $\BB$
is the magnetic field, $\JJ=\nab\times\BB/\mu_0$ the current density,
and $\mu_0$ is the vacuum permeability.
(Here and below, $\bra{...}$ denotes volume averages.)
This is a direct result of magnetic helicity conservation in the
absence of boundaries (B01).
If the turbulence is driven at a scale smaller than the box size $L$,
i.e.\ the forcing wavenumber $k_{\rm f}$ exceeds the box wavenumber
$k_1=2\pi/L$ (so $k_{\rm f}\gg k_1$), it makes sense to use a
two-scale approach. 
We therefore write $\BB=\meanBB+\bb$ and $\JJ=\meanJJ+\jj$, where
the overbar denotes an average field suitably defined over one or
sometimes two periodic coordinate directions, and lower case characters
denote the fluctuations.
The condition of zero current helicity then translates to
\EQ
\bra{\meanJJ\cdot\meanBB}=-\bra{\jj\cdot\bb}
\quad\mbox{(no boundaries)}.
\EN
Together with the assumption that the large and small scale
fields are nearly fully helical and that the sign of the helicity of
the forcing is positive, we have
$\bra{\meanJJ\cdot\meanBB}\approx-k_1\bra{\meanBB^2}/\mu_0$ and
$\bra{\jj\cdot\bb}\approx k_{\rm f}\bra{\bb^2}/\mu_0$.
The important conclusion from this is that, in the steady state, the
amplitude of the mean field exceeds that of the small scale field, with
\EQ
\bra{\meanBB^2}\approx{k_{\rm f}\over k_1}\,\bra{\bb^2}
\quad\mbox{(no boundaries)}.
\EN
Moreover, for large enough magnetic Reynolds numbers the small
scale magnetic energy is in rough equipartition with the turbulent
kinetic energy, i.e.\ $\bra{\bb^2}/\mu_0\approx\bra{\rho\uu^2}$,
but see BB02 for a more accurate estimate
for intermediate values of the magnetic Reynolds number.
In the case considered in B01, where the scale separation ratio
$k_{\rm f}/k_1$ is either 5 or 30, the mean field energy
exceeds the equipartition value by about 5 or 30 -- independent of
the magnetic Reynolds number.

The assumption of full homogeneity, which can only be realized with
triply periodic boundary conditions, was an important ingredient in
arriving at super-equipartition fields.
In this paper we discuss the more general case of non-periodic
boundary conditions.
We use here a mean-field approach together with the dynamical quenching
model (Kleeorin \& Ruzmaikin 1982, Kleeorin et al.\ 1995),
which proved successful in reproducing the homogeneous case
as it was obtained using direct simulations (Field \& Blackman 2002,
BB02, Subramanian 2002).
In the steady state without helicity fluxes, the dynamical quenching model
agrees with a modified catastrophic quenching formula which includes a
term from the current helicity of the large scale field
(Gruzinov \& Diamond 1994, 1995, BB02).

\section{Description of the model}

In the mean field approach we solve the induction equation for the
mean magnetic field $\meanBB$ together with an evolution equation for
the magnetic component of the $\alpha$ effect,
\EQ
{\partial\meanBB\over\partial t}=
\nab\times(\meanUU\times\meanBB+\meanEMF-\eta\meanJJ),
\label{fullset1flux}
\EN
\begin{equation}
{\partial\alpha_{\rm M}\over\partial t}=-2\eta_{\rm t} k_{\rm f}^2\left(
{\meanEMF\cdot\meanBB
+\half k_{\rm f}^{-2}\nab\cdot\meanFF_C^{\rm SS} \over B_{\rm eq}^2}
+{\alpha_{\rm M}\over R_{\rm m}}\right),
\label{fullset2flux}
\end{equation}
where current density is measured in units where $\mu_0=1$,
$\eta$ is the microscopic magnetic diffusivity,
$\eta_{\rm t}$ is the turbulent magnetic diffusivity,
$R_{\rm m}=\eta_{\rm t}/\eta$ is the magnetic Reynolds number,
$\meanEMF$ is the mean electromotive force that includes, among
other terms, a term proportional to $\alpha_{\rm M}\meanBB$.
In the following we adopt the numerical value $k_{\rm f}/k_1=5$
for the scale separation ratio.
The derivation of the $\alpha_{\rm M}$ equation is this normalization can
be found in BB02 without helicity flux and in
Brandenburg \& Sandin (2004, hereafter BS04) with helicity flux.
The connection with $\alpha_{\rm M}=\onethird\tau\overline{\jj\cdot\bb}$
has been accomplished by using the definitions
$B_{\rm eq}^2=\mu_0\rho u_{\rm rms}^2$ and
$\eta_{\rm t}=\onethird\tau u_{\rm rms}^2$, so
$\tau/(3\mu_0\rho)=B_{\rm eq}^2/\eta_{\rm t}$ (see BB02).
In fact, the evolution equation for $\alpha_{\rm M}$ is therefore nothing
else but the evolution equation for the small scale current helicity,
$\overline{\jj\cdot\bb}$.

Indeed, $\overline{\jj\cdot\bb}$ constitutes a possible source of an
$\alpha$ effect, although it occurs usually in conjunction with the
kinetic $\alpha$ effect that, in turn, is proportional to
the negative kinetic helicity, $\overline{\oo\cdot\uu}$,
where $\oo=\nab\times\uu$ is the vorticity.
The two effects together tend to diminish the residual $\alpha$ effect.
We emphasize, however, that this does not need to be the case, and
that even in the absence of kinetic helicity the magnetic
$\alpha$ effect needs to be taken into account.
One example is the case of a decaying helical large scale magnetic field,
where $\overline{\jj\cdot\bb}$ is being generated from the large scale
field.
The associated $\alpha_{\rm M}$ acts as to slow down the decay; see
Yousef et al.\ (2003) for corresponding simulations,
and Blackman \& Field (2004) for related model predictions.

The full expression for $\meanEMF$ can be rather complex.
For the present purpose we restrict ourselves to an expression of the form
\EQ
\meanEMF=\left(\alpha_{\rm K}+\alpha_{\rm M}\right)\meanBB
+\ddelta\times\meanJJ-\eta_{\rm t}\meanJJ,
\label{meanEMFdef}
\EN
where $\alpha_{\rm K}$ is the kinetic $\alpha$ effect,
$\eta_{\rm t}$ is turbulent diffusion, and
$\delta$ can represent both R\"adler's (1969) $\OO\times\meanJJ$ effect
and the $\meanWW\times\meanJJ$ or
shear--current effect of Rogachevskii \& Kleeorin (2003, 2004).
(Here $\meanWW = \nab \times \meanUU$ is the vorticity of the mean flow.)
The importance of shear, $S$ ($=\partial U_y/\partial x$ in some of the
first cases reported blow), and of the two turbulent dynamo effects,
$\alpha$ and $\delta$, is quantified in terms of the non-dimensional numbers
\EQ
C_S={S\over\eta_{\rm t}k_1^2},\quad
C_\alpha={\alpha_{\rm K}\over\eta_{\rm t}k_1},\quad
C_\delta={\delta\over\eta_{\rm t}}.
\EN
In the following we restrict ourselves to cases where either
$C_\alpha$ or $C_\delta$ are different from zero.

The general importance of helicity fluxes has been identified by
Blackman \& Field (2000a,b) and Kleeorin et al.\ (2000).
Here we use a generalized from of the current helicity flux of
Vishniac \& Cho (2001) flux, as derived by Subramanian \& Brandenburg (2004),
\EQ
\meanFF^{\rm SS}_{C\,i}=\phi_{ijk}\meanB_j\meanB_k.
\label{helflux1}
\EN
Under the assumption that $\nab\cdot\meanUU=0$, we show in Appendix~A that
\EQ
\phi_{ijk}=C_{\rm VC}\,\epsilon_{ijl}\overline{\sf S}_{lk},
\label{helflux2}
\EN
where $\overline{\sf S}_{lk}=\half(\meanU_{l,k}+\meanU_{k,l})$ is the
mean rate of strain tensor and $C_{\rm VC}$ is a non-dimensional coefficient
that is of order unity (see Appendix~A).
In the following we consider $C_{\rm VC}$ as a free parameter.
It turns out that  there is a critical value, $C_{\rm VC}^*$, above
which there is runaway growth that can only be stopped by adding
an extra quenching term.
One possibility is to consider an algebraic quenching of the total
$\alpha$ effect ($\alpha=\alpha_{\rm K}+\alpha_{\rm M}$).
Here we use a rough and qualitative approximation to the full expressions
of Kleeorin \& Rogachevskii (2002) by using
\EQ
\alpha=\alpha_0/\left(1+g_\alpha\meanBB^2/B_{\rm eq}^2\right),
\label{algebraic}
\EN
where we choose $g_\alpha=3$ as a good approximation to the
full expression (see BB02 for a discussion in similar context).
For a completely independent and purely numerical verification of
algebraic and non-$R_{\rm m}$ dependent quenching of $\alpha_{\rm K}$
and $\alpha_{\rm M}$ see Brandenburg \& Subramanian (2005b).

We expect the critical value $C_{\rm VC}^*$ to decrease with increasing
value of $C_S$.
However, since $C_{\rm VC}$ should normally be fixed by physical
considerations (which are uncertain), the possibility of a critical
state translates to a critical value of $C_S$, above which ``strong''
(or ``runaway'') dynamo action is possible.
This effect is similar to the $\meanWW\times\meanJJ$ effect
in that it only requires nonhelical turbulence and shear.
However, we will show that, unless there is also current helicity
flux above a certain threshold, the field generated by the
$\meanWW\times\meanJJ$ effect effect alone is weak when there
are boundaries and when the magnetic Reynolds number is large.
For the same reason, also $\alpha$ effect dynamos produce only week
fields, unless the current helicity flux exceeds a certain threshold.

Once the current helicity flux is supercritical, it is important
to make sure that $\alpha_{\rm M}$ is spatially smooth.
This is accomplished by adding a small diffusion term of the form
$\kappa_\alpha\nabla^2\alpha_{\rm M}$ to the right hand side of
\Eq{fullset2flux}.
(Typical values considered below are $\kappa_\alpha=0.02\nu_{\rm t}$.)

We consider both one-dimensional and two-dimensional models.
In both cases we allow for the possibility of shear.
In the one-dimensional case ($-L/2<z<L/2$)
we allow for a linear shear flow of the form $\meanUU=(0,Sx,0)$,
so the mean field dynamo equation is given by
\EQ
{\partial\meanB_x\over\partial t}=
-{\partial\meanemf_y\over\partial z}
+\eta{\partial^2\meanB_x\over\partial z^2},
\label{dBxdz}
\EN
\EQ
{\partial\meanB_y\over\partial t}=S\meanB_x+{\partial\meanemf_x\over\partial z}
+\eta{\partial^2\meanB_y\over\partial z^2},
\label{dBydz}
\EN
and the current helicity flux is given by
\EQ
\zz\cdot\meanF^{\rm SS}_{C}=\half C_{\rm VC}\, S(\meanB_x^2-\meanB_y^2),
\EN
where $\zz$ is the unit vector in the $z$ direction.
Note that, according to this formula, assuming $S>0$,
and because $\meanB_x^2<\meanB_y^2$ in such a shear flow,
negative current helicity flows in the positive $z$ direction.
This is also the direction of the dynamo wave for $\alpha_{\rm K}>0$.
[We recall that for $\alpha_{\rm K}>0$, negative current helicity must
be lost to alleviate catastrophic quenching (Brandenburg et al.\ 2002).]
We assume va\-cu\-um boundary conditions,
\EQ
\meanB_x=\meanB_y=0\quad\mbox{(on $z=\pm L/2$)},
\label{bc}
\EN
which implies that $\nnn\cdot\meanFF^{\rm SS}_{C}=0$ on the boundaries.
This property can be regarded as an unfortunate shortcoming of the present
model, because, although current helicity can efficiently be transported
to the vicinity of the boundary, it is actually unable to leave the domain.
On the other hand, this type of boundary condition was also used in
the simulations presented in
Brandenburg (2005, hereafter referred to as B05) where the absence of
closed boundaries clearly did allow for a significantly enhanced final
field strength (see also Brandenburg et al.\ 2005).
(It is also possible that in the simulations current helicity of the small
scale field got lost because of numerical dissipation on the boundaries.)

In order to allow for a finite current helicity flux on the boundaries
we also compare with the case of an extrapolating boundary condition,
\EQ
\epsilon z{\partial\meanB_i\over\partial z}+\meanB_i=0
\quad\mbox{(on $z=\pm L/2$ for $i=x,y$)}.
\label{modified_bc}
\EN
Thus, if $\epsilon=0$ we recover the standard vacuum boundary
condition, $\meanB_i=0$.
For $\epsilon>0$, the slope of $\meanB_i(z)$ is such that $\meanB_i$
would vanish outside the domain on fiducial reference points,
$z=\pm(1+\epsilon)L/2$.

In the two-dimensional simulations we consider two different cases.
In the first case we use \Eq{bc} in the vertical direction and periodic
boundary conditions in the horizontal.
In the second case we use perfectly conducting and pseudo-vacuum
boundary conditions on the four boundaries in a meridional
cross-section, just like in the corresponding direct simulations (BS04).
The pseudo-vacuum boundary conditions are applied on what would correspond
in the sun to the outer surface and the equatorial plane.

In all cases the initial magnetic field is a random field of sufficiently
small amplitude, so that the nonlinear solutions grow out of the linear one.
We have not made a serious attempt to search for solutions that only
exist as finite amplitude solutions.
Furthermore, we assume that initially $\alpha_{\rm M}=0$, which is sensible
if one starts with weak initial fields.
Again, we have not made a systematic search for finite amplitude solutions
that might only be accessible with finite initial values of $\alpha_{\rm M}$.

\section{Results}

\subsection{Reference case with no shear}

We begin with the case of an $\alpha^2$ dynamo where $S=0$,
so there is no shear and hence no helicity flux.
The dynamo is excited for $\alpha>\eta_{\rm T}k_1$.
A possible solution that is marginally excited and satisfies the
boundary condition \eq{bc} is given by
\EQ
\meanBB(z)=\pmatrix{1+\cos k_1z\cr-\sin k_1z\cr0};
\EN
see Meinel \& Brandenburg (1990) for the more general case
of non-marginally excited (but still only kinematic) solutions.

\begin{figure}[t!]\begin{center}
\includegraphics[width=\columnwidth]{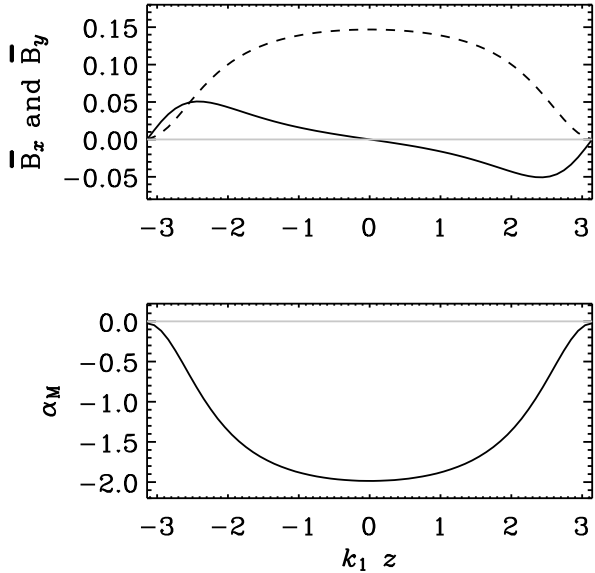}
\end{center}\caption[]{
Field structure for an $\alpha^2$ dynamo
(solid line $\meanB_x/B_{\rm eq}$, dashed line $\meanB_y/B_{\rm eq}$)
together with $\alpha_{\rm M}$, for
$C_\alpha=3$, and $R_{\rm m}=10^2$.
}\label{pb_alpha}\end{figure}

\begin{figure}[t!]\centering
\includegraphics[width=\columnwidth]{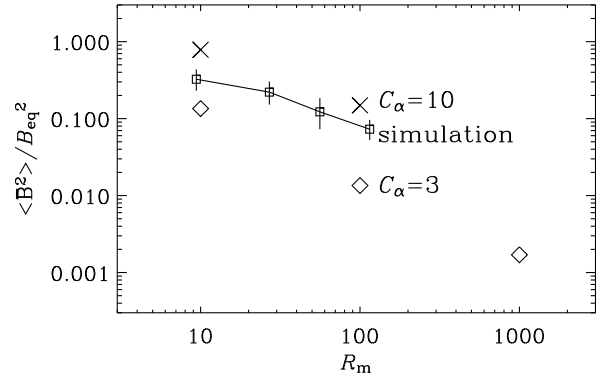}\caption{
Dependence of $\bra{\meanBB^2}/B_{\rm eq}^2$ on the magnetic
Reynolds number for a run with open boundary conditions and no shear
both for the mean field model and the direct simulation (squares
connected by a line, with approximate error bars).
The diamonds and crosses refer to mean field models where
$C_\alpha$ is 3 and 10, respectively.
}\label{pbmrm}\end{figure}

We have solved \Eqs{dBxdz}{dBydz} numerically using a third-order
Runge-Kutta time stepping scheme and a sixth-order finite difference scheme.
An example of a solution is shown in \Fig{pb_alpha}.
The results are displayed in \Tab{Tres1} and \Fig{pbmrm}.
Both simulation data and mean field models are roughly
compatible with the relation
\EQ
\bra{\meanBB^2}/B_{\rm eq}^2\propto R_{\rm m}^{-1}
\quad\mbox{(with boundaries)},
\label{B2scalingRm}
\EN
that was first found analytically by Gruzinov \& Diamond (1995) for the
same boundary conditions \eq{bc}.
The simulation data shown in \Fig{pbmrm}
supersede earlier results of Brandenburg \& Dobler (2001) at
lower resolution and smaller values of $R_{\rm m}$ where the scaling
seemed compatible with $\meanBB^2\sim R_{\rm m}^{-1/2}$.
However, in view of the new results
this must now be regarded as an artifact of insufficient dynamical range,
so the correct scaling is given by \Eq{B2scalingRm}.
Furthermore, the simulation results give agreement with the mean field
model if the dynamo number, $C_\alpha$, is somewhere between 3 and 10.
This appears compatible with the fact that the scale separation ratio
in the simulation is $k_{\rm f}/k_1=5$, and that this ratio gives a
good estimate of $C_\alpha$; see BB02.

\begin{table}[b!]\caption{   
Mean squared field strength of dynamically quenched $\alpha^2$ dynamos.
For time dependent solutions $\meanBB^2$ is also averaged in time, and
the frequency of the solution is given.
Note in particular the inverse proportionality between
$\bra{\meanBB^2}$ and $R_{\rm m}$ in the first three rows
(marked by asterisks).
}\vspace{12pt}\centerline{\begin{tabular}{clccccccc}
 & $R_{\rm m}$ &  $C_\alpha$  
& $\bra{\meanBB^2}/B_{\rm eq}^2$
& $\omega/(\eta_{\rm T}k_1^2)$ \\
\hline
$*$ & $10^1$ & 3 & $1.35\times10^{-1}$ &   0  \\  %OK
$*$ & $10^2$ & 3 & $1.35\times10^{-2}$ &   0  \\  %OK
$*$ & $10^3$ & 3 & $1.74\times10^{-3}$ & 0.28 \\  %1d/64a
  & $10^1$ & 10& $7.85\times10^{-1}$ & 0.85 \\  %corrected
  & $10^2$ & 10& $1.49\times10^{-1}$ & 0.64 \\  %corrected
\label{Tres1}\end{tabular}}\end{table}

Returning now to the description of the mean field calculations, we note that
at large values of $R_{\rm m}$ the system shows relaxation oscillations
where the sign of the field does not necessarily change
(so the period of the field agrees with the period of the rms value).
The frequency given in the table is $\omega=2\pi/T_{\rm period}$.

\begin{figure*}[t!]\begin{center}
\includegraphics[width=\textwidth]{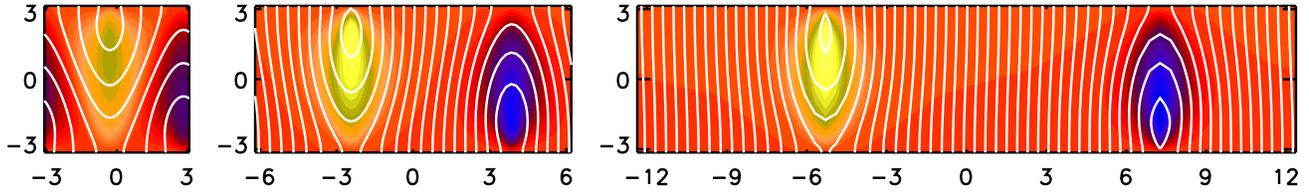}
\end{center}\caption[]{
Field lines superimposed on a color/gray scale representation of
the normal field component for an $\alpha^2$ dynamo in two dimensions
in the $xz$ plane with $C_\alpha=3$ and three different aspect ratios
(1, 2, and 4).
Note that there remain only two cells even for large aspect ratios.
}\label{paspect}\end{figure*}

\begin{table}[b!]\caption{   
Same as in \Tab{Tres1}, but for two-dimensional calculations for different
values of $R_{\rm m}$ and different aspect ratios $L_x/L_z$.
The inverse proportionality between $\bra{\meanBB^2}$ and $R_{\rm m}$
can best be seen from the last three rows (marked by asterisks).
The effect of changing the aspect ratio can be seen by inspecting
the first two and the fourth row.
}\vspace{12pt}\centerline{\begin{tabular}{clccccccc}
 & $R_{\rm m}$ &  $C_\alpha$ & $L_x/L_z$
& $\bra{\meanBB^2}/B_{\rm eq}^2$
& $\omega/(\eta_{\rm T}k_1^2)$ \\
\hline
 & $10^2$ & 3 & 1 & $1.06\times10^{-1}$ &   0  \\  %64b
 & $10^2$ & 3 & 2 & $2.30\times10^{-1}$ &   0  \\  %64c
$*$ & $10^1$ & 3 & 4 & $1.25\times10^{-0}$ &   0  \\  %64d_Rm10
$*$ & $10^2$ & 3 & 4 & $2.53\times10^{-1}$ &   0  \\  %64d_Rm1e2
$*$ & $10^3$ & 3 & 4 & $2.71\times10^{-2}$ &   0  \\  %64d_Rm1e3
\label{Tres1_2D}\end{tabular}}\end{table}

The fact that the saturation field strength decreases with increasing
magnetic Reynolds number is bad news for astrophysical applications.
However, this result is in agreement with simulations, lending thereby
support to the applicability of mean field theory.

The situation may change by allowing the field to extend also in
one of the perpendicular directions (say the $x$ direction), because
then the field may have its main variation in this new direction.
In particular, if the field were to display Beltrami-like behavior
in that direction, it might be more similar to the case of triply
periodic boundary conditions.
However, as can be seen from \Tab{Tres1_2D}, this does not seem to
be the case.
Even at large aspect ratios the wavelength of the field in the
$x$ direction remains of the order of the extent of the domain
in that direction (\Fig{paspect}), and the field amplitude still decreases
inversely proportional to the magnetic Reynolds number (\Tab{Tres1_2D}).

\subsection{Solutions with uniform shear}

For positive values of $\alpha_{\rm K}$ and positive values of $S$ there are
dynamo waves traveling in the positive $z$ direction.
This is also the direction in which the flux of negative current helicity
is pointing.
Note that the saturation magnetic field strength decreases with increasing
magnetic Reynolds number in the same way as before (\Tab{Ttimescale}).
The modified boundary condition \eq{modified_bc} with $\epsilon\neq0$ does
lead to an increase of the saturation field strength for $R_{\rm m}=10^2$,
but not for $R_{\rm m}=10^3$.
This behavior is not altered by the presence of helicity fluxes,
i.e.\ changing the value of $C_{\rm VC}$ from 0 to 0.2 has only a small effect.

\begin{figure}[t!]\begin{center}
\includegraphics[width=\columnwidth]{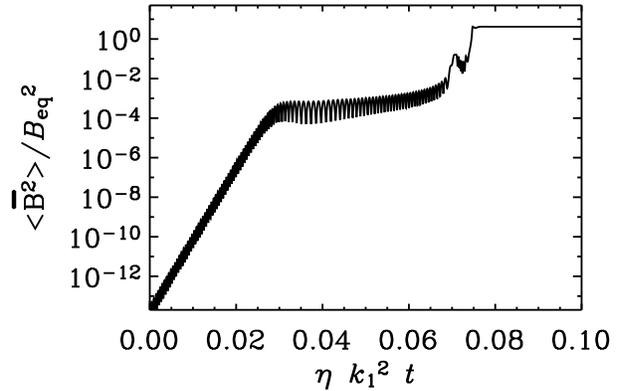}
\end{center}\caption[]{
Evolution of $\bra{\meanBB^2}/B_{\rm eq}^2$ for a model with
$C_{\rm VC}=1$ and additional algebraic quenching with $g_\alpha=3$,
$\kappa_\alpha=0.02\eta_{\rm t}$,
$C_S=10$, $C_\alpha=0.2$, and $R_{\rm m}=10^4$.
Note that the abscissa is scaled in resistive time units.
}\label{pg3}\end{figure}

For $C_S=10$, the critical value for runaway dynamo action is
$C_{\rm VC}^*=0.15$, but it decreases with increasing shear
(e.g.\ for $C_S=20$ we have $C_{\rm VC}^*=0.036$.)
This runaway growth may be a possible solution to the quenching problem
in that it allows the solution to continue growing until
it is saturated by other effects such as the usual $\alpha$ quenching
that is independent of $R_{\rm m}$; see \Eq{algebraic}.
Note that when runaway occurs, the magnetic field increases sharply
until it reaches a new saturation value close to equipartition;
see \Fig{pg3}, where $C_{\rm VC}=1$ has been chosen.
However, the magnetic field is then no longer oscillatory.
Obviously, the algebraic quenching adopted here is not realistic,
but it does at least illustrate the point that when $C_{\rm VC}$
exceeds a certain critical value, there is runaway that could
potentially be contained by having additional quenching terms.

\begin{table}[b!]\caption{   
Mean squared field strength of dynamically quenched $\alpha^2\Omega$ dynamos.
For time dependent solutions $\meanBB^2$ is also averaged in time, and
the frequency of the solution is given.
For all calculations we used $64$ meshpoints.
For $C_\alpha=0.2$ and $C_S=10$ the kinematic growth rate is 0.08.
The inverse proportionality between $\bra{\meanBB^2}$ and $R_{\rm m}$
can best be seen from the first three rows (marked by asterisks).
Changing the sign of $C_S$ has no effect on the saturation field strengths,
regardless of the value of $C_{\rm VC}$.
}\vspace{12pt}\centerline{\begin{tabular}{clccccccc}
& $R_{\rm m}$ & $\epsilon$ & $C_{\rm VC}$
& $\bra{\meanBB^2}/B_{\rm eq}^2$ & $\omega/(\eta_{\rm T}k_1^2)$ \\
\hline
$*$ & $10^1$ & 0 &  0  & $1.52\times10^{-2}$ & 0.53 \\ %64d2
$*$ & $10^2$ & 0 &  0  & $1.49\times10^{-3}$ & 0.56 \\ %64b2
$*$ & $10^3$ & 0 &  0  & $2.49\times10^{-4}$ & 0.54 \\ %64c
    & $10^1$ & 0 & 0.2 & $3.18\times10^{-2}$ & 0.44 \\ %64d
    & $10^2$ & 0 & 0.2 & $1.57\times10^{-3}$ & 0.57 \\ %64b
    & $10^1$ & 0&$-0.2$& $1.06\times10^{-2}$ & 0.56 \\ %64d3
    & $10^2$ & 0&$-0.2$& $1.35\times10^{-3}$ & 0.57 \\ %64b3
\hline
    & $10^1$ &0.2&  0  & $3.58\times10^{-3}$ & 0.59 \\ %idl
    & $10^2$ &0.2&  0  & $7.73\times10^{-2}$ & 0.36 \\ %idl
    & $10^1$ &0.2& 0.2 & $4.31\times10^{-3}$ & 0.58 \\ %idl
    & $10^2$ &0.2& 0.2 & $1.45\times10^{-2}$ & 0.22 \\ %idl
    & $10^3$ &0.2& 0.2 & $4.26\times10^{-4}$ & 0.47 \\ %idl
\label{Ttimescale}\end{tabular}}\end{table}

\begin{figure}[t!]\begin{center}
\includegraphics[width=\columnwidth]{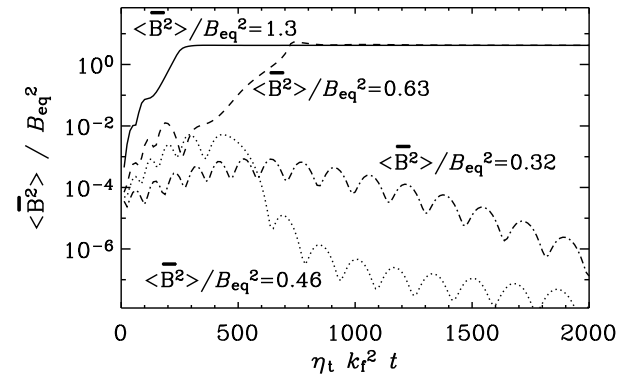}
\end{center}\caption[]{
Evolution of $\bra{\meanBB^2}/B_{\rm eq}^2$ of dynamo with no kinetic
helicity $C_\alpha=0$, just shear with $C_S=10$ and a supercritical
current helicity flux with $C_{\rm VC}=1$, and different initial
field strengths.
(Because the initial field is random, much of the initial energy is
quickly lost by dissipation in the high wavenumbers, which cannot
be seen in the plot.)
In all cases $R_{\rm m}=10^4$.
Note that the abscissa is scaled in dynamical time units.
}\label{pcompsat_vc}\end{figure}

As was already anticipated by Vishniac \& Cho (2001), a supercritical
current helicity flux could by itself also drive a mean field dynamo.
This mechanism requires a finite amplitude initial magnetic field
to get started; see \Fig{pcompsat_vc}.
With unsuitable initial conditions the dynamo may therefore not
get started and one might miss it.
It is also possible that, even though the expected helicity flux is
present and of the right kind, its strength remains subcritical
(Arlt \& Brandenburg 2001).

\begin{figure}[t!]\begin{center}
\includegraphics[width=\columnwidth]{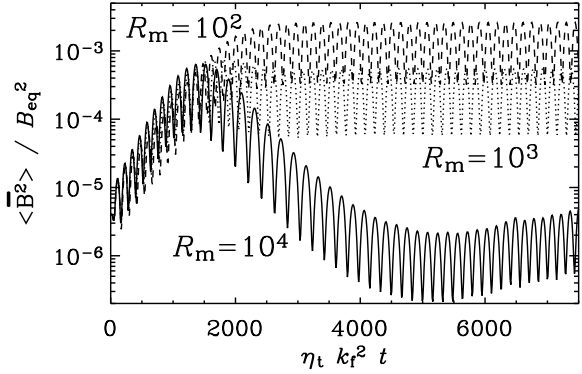}
\end{center}\caption[]{
Evolution of $\bra{\meanBB^2}/B_{\rm eq}^2$ for the
$\alpha^2\Omega$ dynamo with
$C_\alpha=0.2$, $C_S=10$, $C_{\rm VC}=0.2$, and different values
of $R_{\rm m}$.
Note that the abscissa is scaled in dynamical time units.
}\label{pcompsat}\end{figure}

It is interesting to note that much of the late kinematic
phase does not depend on the value of $R_{\rm m}$; see \Fig{pcompsat}.
In fact, during the kinematic phase the runs with larger values of
$R_{\rm m}$ have slightly larger magnetic energies.
Within a time scale that is independent of $R_{\rm m}$ [here 2000
dynamical time units, $(\eta_{\rm t}k_{\rm f}^2)^{-1}$] the large
scale magnetic energy reaches a significant field strength whose peak
value is roughly independent of the magnetic Reynolds number.
If one discards the subsequent decline of the magnetic energy, this
result would be similar to that in the homogeneous case.
In the weakly supercritical case,
the cycle frequency does not decrease as nonlinearity becomes important.
This is indeed an important feature of the dynamical quenching model that
distinguishes it from the catastrophic quenching hypothesis (BB02).
The run presented in \Fig{pcompsat} is for a finite current helicity
flux ($C_{\rm VC}=0.2$), but the qualitative form of this plot is actually
independent of helicity fluxes and can also be obtained for $C_{\rm VC}=0$,
for example.

\begin{figure}[t!]\begin{center}
\includegraphics[width=\columnwidth]{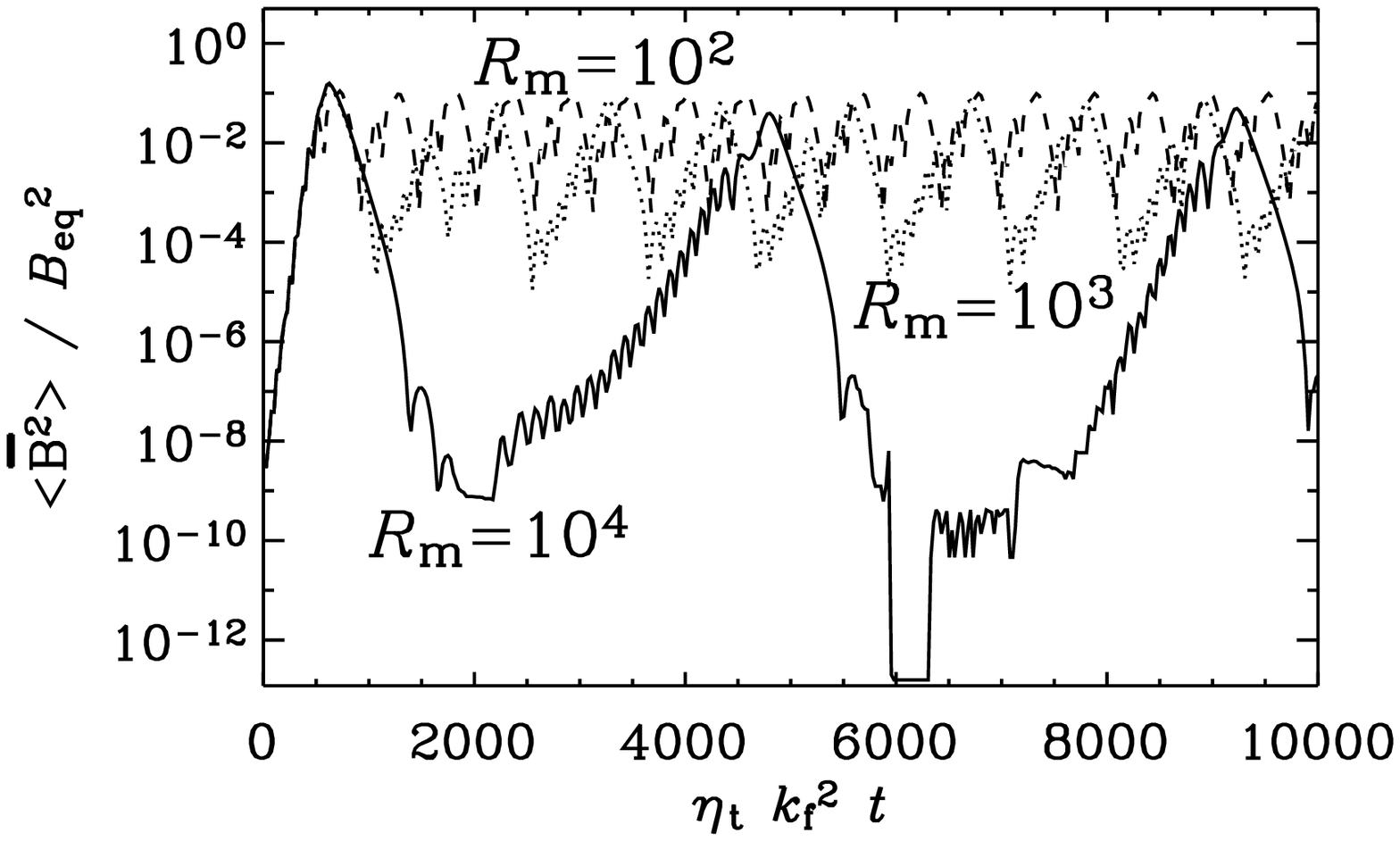}
\end{center}\caption[]{
Evolution of $\bra{\meanBB^2}/B_{\rm eq}^2$ for the
$\alpha^2\Omega$ dynamo with
$C_\alpha=0.2$, $C_S=30$, $C_{\rm VC}=0$, and different values
of $R_{\rm m}$. 
The line styles are as in \Fig{pcompsat} above.
}\label{pcompsat2}\end{figure}

If the dynamo number is increased further to be highly supercritical
then the peak value reached by the large scale field, at the
end of the late kinematic stage, increases even further;
see \Fig{pcompsat2}. This is again as anticipated by BB02 and 
Subramanian (2002). However, what was not anticipated are the subsequent 
strong dips in the mean field energy. It appears that the growth of 
$\alpha_{\rm M}$ and the subsequent decrease of $\alpha$, takes the dynamo
below criticality. The mean field then decays until the microscopic
diffusivity term [last term in \Eq{fullset2flux}] causes
$\alpha_{\rm M}$ to decay on a resistive timescale, and the net 
$\alpha$ to again increase, such that the dynamo becomes 
supercritical again. This qualitatively accounts for the long term
oscillations of the mean field energy seen in \Fig{pcompsat2},
whose period clearly increases with $R_{\rm m}$. It however implies
that, on the average, the mean field energy again decreases
with increasing $R_{\rm m}$, when there is no flux.

\subsection{Shear-current effect}

The issue of catastrophic quenching is often associated with the
$\alpha$ effect alone, and it is implied that other large scale
dynamo effects may not have this problem.
This is however not true.
The main problem is quite generally associated with the helical
nature of the large scale magnetic field.
As is well known, the $\meanWW\times\meanJJ$ effect can exist
even without kinetic helicity and just shear alone
(Rogachevskii \& Kleeorin 2003, 2004).
In the presence of closed boundaries, the current helicity of the
large scale field results in a corresponding contribution from the
small scale field, which affects the resulting electromotive force.
This can be seen quite generally by considering the stationary
limit of \Eq{fullset2flux} for the case $C_{\rm VC}=0$,
which yields $\alpha_{\rm M}=-R_{\rm m}\meanEMF\cdot\meanBB/B_{\rm eq}^2$,
\EQ
\meanEMF\cdot\meanBB={\meanEMF_0\cdot\meanBB\over
1+R_{\rm m}\meanBB^2/B_{\rm eq}^2}\quad\mbox{(steady state)}.
\EN
Here we have defined the unquenched electromotive force
\EQ
\meanEMF_0=\alpha_{\rm K}\meanBB+\ddelta\times\meanJJ-\eta_{\rm t}\meanJJ,
\EN
where the $\alpha_{\rm M}$ term is absent compared with \Eq{meanEMFdef}.
To illustrate the catastrophic quenching of the $\meanWW\times\meanJJ$ effect
we set $\alpha_{\rm K}=0$ and $\ddelta=(0,0,\delta)^{\rm T}$.

\begin{figure}[t!]\begin{center}
\includegraphics[width=\columnwidth]{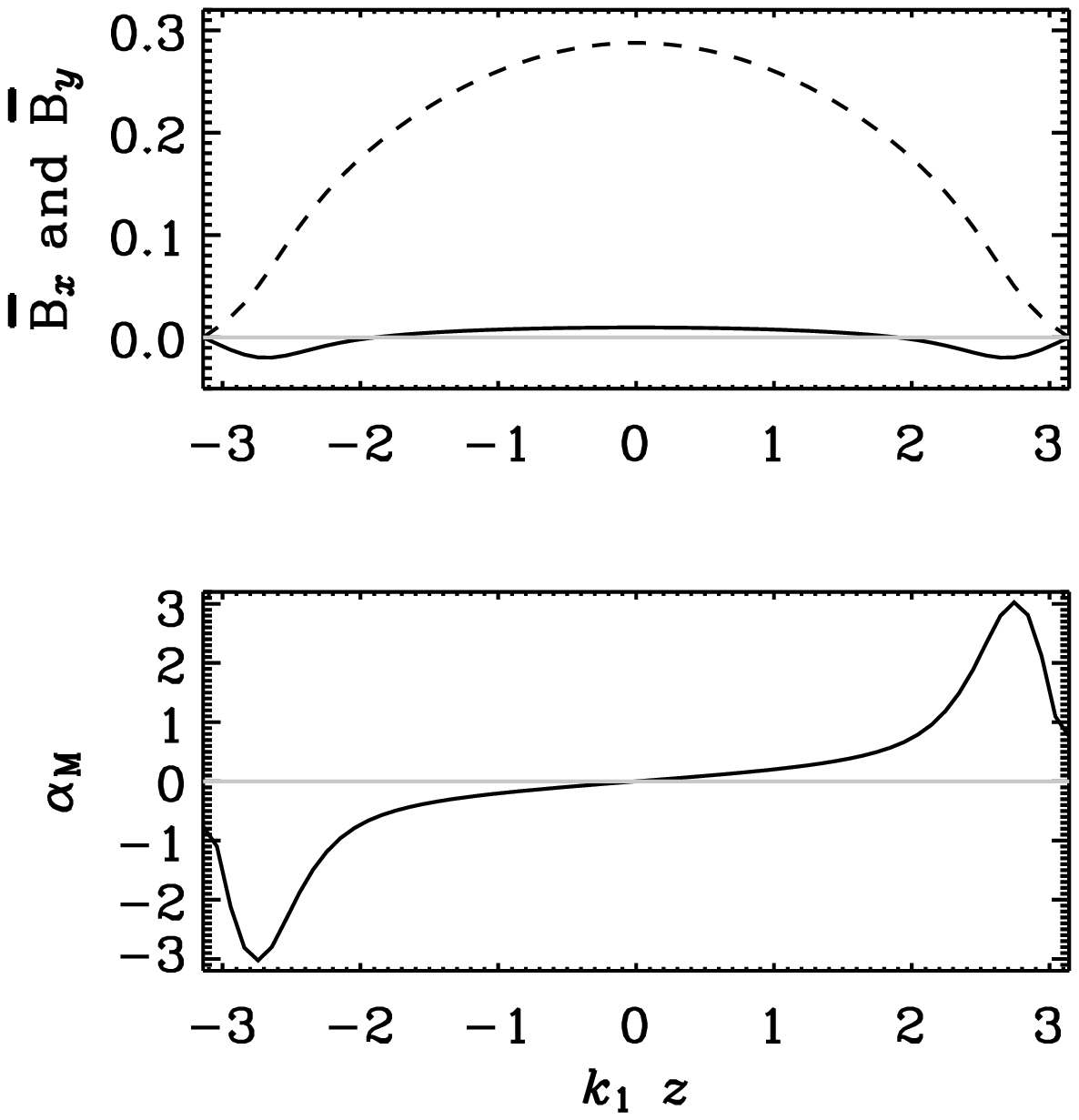}
\end{center}\caption[]{
Field structure for a $\ddelta\times\meanJJ$ dynamo
(solid line $\meanB_x/B_{\rm eq}$, dashed line $\meanB_y/B_{\rm eq}$)
together with $\alpha_{\rm M}$, for
$C_\delta=-1$, $C_S=2$, $C_{\rm VC}=2$, and $R_{\rm m}=10^3$.
}\label{pb_delta}\end{figure}

As is already clear from linear theory, a $\ddelta\times\meanJJ$ effect
can only produce self-excited solutions if $\delta/S<0$
(e.g.\ Brandenburg \& Subramanian 2005a), where the shear associated with
$\meanWW$ is responsible for stretching the poloidal field into toroidal.
The $\ddelta\times\meanJJ$ effect can also convert poloidal field into toroidal
field, in addition to converting toroidal into poloidal, but this effect
alone would not produce energy in the mean magnetic field.
This is why shear is necessary.
[We note, however, that there is currently a controversy regarding the
expected sign of the $\ddelta\times\meanJJ$ effect; see
R\"udiger \& Kitchatinov (2005).]

The field geometry is shown in \Fig{pb_delta}; it resembles
that of an $\alpha^2$ dynamo (cf.\ \Fig{pb_alpha}), except that
now both $B_x$ and $B_y$ are symmetric about the midplane, and
$\alpha_{\rm M}$ is antisymmetric about $z=0$.
(The symmetry property of $\alpha_{\rm M}$ is identical to that
of $\meanEMF\cdot\meanBB$, where one contribution is
$(\ddelta\times\meanJJ)\cdot\meanBB=\half\ddelta\cdot\nab\meanBB^2$.)
In \Tab{Tres2} we present the results for different values of $R_{\rm m}$.
Again, we see quite unambiguously that, for a fixed value of $S$, the resulting
field strength decreases with increasing $R_{\rm m}$, just like in all
previous cases.

\begin{table}[t!]\caption{
Mean squared field strength of dynamically quenched $\delta^2\Omega$ dynamos.
For all calculations we used $64$ meshpoints.
The inverse proportionality between $\bra{\meanBB^2}$ and $R_{\rm m}$
can again be seen from the last two rows (marked by asterisks).
}\vspace{12pt}\centerline{\begin{tabular}{lccccccc}
& $R_{\rm m}$ & $C_\delta$ & $C_S$ & $C_{\rm VC}$
& $\bra{\meanBB^2}/B_{\rm eq}^2$ & $\omega/(\eta_{\rm T}k_1^2)$ \\
\hline
    & $10^2$ & $-1.0$& 2 & 0 & $1.48\times10^{-1}$ &  0  \\
    & $10^2$ & $-1.0$& 2 & 1 & $2.21\times10^{-1}$ &  0  \\
$*$ & $10^2$ & $-1.0$& 2 & 2 & $4.24\times10^{-1}$ &  0  \\
$*$ & $10^3$ & $-1.0$& 2 & 2 & $4.39\times10^{-2}$ &  0  \\
$*$ & $10^4$ & $-1.0$& 2 & 2 & $4.16\times10^{-3}$ &  0  \\
\label{Tres2}\end{tabular}}\end{table}

\begin{table}[b!]\caption{
Mean squared field strength of dynamically quenched $\alpha^2\Omega$ dynamos
with a solar-like shear profile.
All solutions are time dependent, so $\meanBB^2$ is also averaged in time,
and the frequency of the field (not its energy) is given.
The inverse proportionality between $\bra{\meanBB^2}$ and $R_{\rm m}$
can be seen from the first three rows (marked by asterisks).
For all calculations we used $128^2$ meshpoints using
$C_\alpha=3$ and $C_S=10^3$.
}\vspace{12pt}\centerline{\begin{tabular}{lccccccc}
    & $R_{\rm m}$ & $C_\alpha$ & $C_S$ & $C_{\rm VC}$
& $\bra{\meanBB^2}/B_{\rm eq}^2$ \\
\hline
$*$ & $10^1$ & 3 & $10^3$ & 0.1 & $9.4\times10^{-2}$ &      \\ %VC64a_Rm1e1
$*$ & $10^2$ & 3 & $10^3$ & 0.1 & $9.2\times10^{-3}$ &      \\ %VC128a_Rm1e2
$*$ & $10^3$ & 3 & $10^3$ & 0.1 & $8.4\times10^{-4}$ &      \\ %VC64a_Rm1e3
\hline
    & $10^2$ & 3 & $10^3$ &0.001& $2.96\times10^{-1}$ &      \\ %VC128c_Rm1e2
    & $10^2$ & 3 & $10^3$ & 0.01& $4.98\times10^{-2}$ &      \\ %VC128a_Rm1e2
    & $10^2$ & 3 & $10^3$ & 0.1 & $9.22\times10^{-3}$ &      \\ %VC128a_Rm1e2
\label{SolarLike}\end{tabular}}\end{table}

\subsection{Solar-like shear}

Finally, we consider the more complex flow geometry employed by
BS04 and B05 in an attempt to
approximate the differential rotation profile in lower latitudes
of the sun (cf.\ Figs~1 and 2 of BS04).
In the simulations presented in these two papers the turbulence was
forced in such a way that it has either positive, negative, or zero kinetic
helicity.
The latter case would correspond to no kinetic $\alpha$ effect, but the
Vishniac \& Cho mechanism and the
$\meanWW\times\meanJJ$ effect may still provide a possible explanation
for the large scale field that is actually generated in such a simulation
(B05).

\begin{figure}[t!]\begin{center}
\includegraphics[width=\columnwidth]{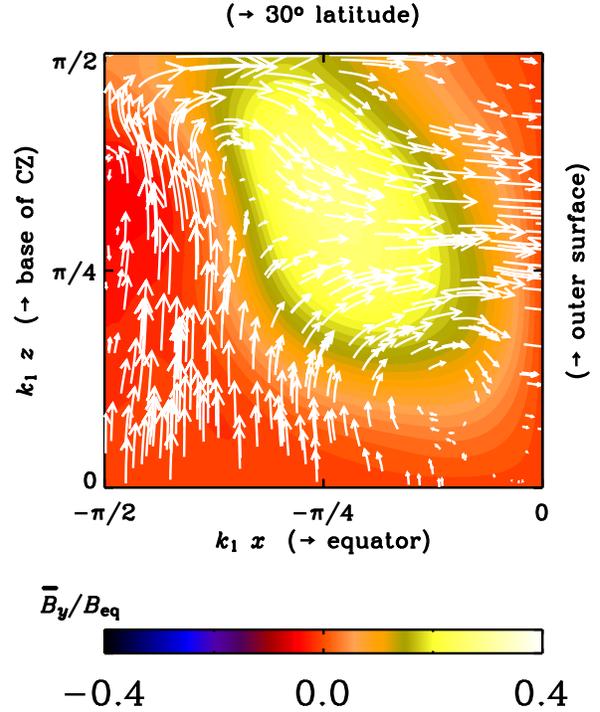}
\end{center}\caption[]{
Snapshot showing the field structure for an $\alpha^2\Omega$ dynamo
with solar-like shear, $C_\alpha=3$, $C_S=10^3$, $C_{\rm VC}=10^{-3}$,
and $R_{\rm m}=10^2$.
The solution remains highly time-dependent.
}\label{pb}\end{figure}

We use here the solar-like shear profile of BS04 that is given by
\EQ
\meanUU=(S/k_1)(0,\cos k_1 x\,\cos k_1 z,0)^{\rm T}
\EN
in the domain $-\pi/2\leq k_1x\leq0$, $0\leq k_1z\leq\pi/2$.
As discussed in BS04, $k_1 x=-\pi/2$ corresponds to the bottom of the
convection zone, where the toroidal flow is constant and approximately
equal to that at 30 degrees latitude (corresponding to $k_1z=\pi/2$;
the surfaces at $z=0$ and $x=0$ correspond to equator and outer surface,
respectively, and permit current helicity fluxes.

Length is measured in units of the inverse basic wavenumber of the domain,
$k_1^{-1}$.
Hereafter we assume $k_1=1$.
The rate of strain matrix is then given by
\EQ
\overline{\SSSS}=-\half S\pmatrix{
0              & \sin x\,\cos z & 0              \cr
\sin x\,\cos z & 0              & \cos x\,\sin z \cr
0              & \cos x\,\sin z & 0              \cr
},
\EN
so the divergence of the current helicity flux is, using
\Eqs{helflux1}{helflux2}, given by
\EQA
{\nab\cdot\meanFF_C^{\rm SS}\over\half S C_{\rm VC}}
&\!=\!&
-\cos x\,\sin z\left[
   {\partial\over\partial x}\left(\meanB_y^2-\meanB_z^2\right)
  +{\partial\over\partial z}\left(\meanB_x\meanB_z\right)\right]
\nonumber \\
&&+\sin x\,\cos z\left[
   {\partial\over\partial z}\left(\meanB_y^2-\meanB_x^2\right)
  +{\partial\over\partial x}\left(\meanB_x\meanB_z\right)\right]
\nonumber \\
&&+\sin x\,\sin z\left(\meanB_x^2-\meanB_z^2\right).
\ENA
The corresponding value of $\nab\cdot\meanFF_C^{\rm SS}$ is used in
\Eq{fullset2flux}, and the dynamo equation \eq{fullset1flux} is solved
subject to the same boundary conditions used in BS04.

The field turns out to be highly irregular in time.
An example of a snapshot at an arbitrarily chosen moment in time
is given in \Fig{pb}.
The magnetic field looks roughly similar to that found by averaging
the results of direct simulations; see Fig.~6 of B05
or Fig.~18 of Brandenburg et al.\ (2005).

In \Tab{SolarLike} we give the time and volume averaged values of the
squared mean field for different values of $R_{\rm m}$ and different
values of $C_{\rm VC}$.
Again, note that the energy of the mean field decreases inversely
proportional to the magnetic Reynolds number.
More surprisingly, increasing the value of $C_{\rm VC}$ has an adverse effect
on the saturation field amplitude.

\section{Conclusions}

The present investigations still leave us with a puzzle.
On the one hand the dynamically quenched mean field models
yield invariably a resistively quenched saturation amplitude
of the mean field -- regardless of details of the boundary conditions,
the presence of shear, or the nature of the dynamo effect.
(In the absence of shear and just $\alpha^2$ dynamo action, this result
of mean field theory is well confirmed by turbulence simulations.)
On the other hand, simulations with open boundary conditions
(B05) have shown a clear difference compared with
the case of closed boundaries.
The reason for this discrepancy remains unclear at this point.
It is however possible that $C_{\rm VC}$ is simply large enough,
so there is a runaway dynamo effect driving large scale fields
by the Vishniac \& Cho mechanism.
A possible argument against this explanation, is that in B05,
the dynamo did not resemble threshold behavior.

Even though for subcritical helicity fluxes
the saturation field strength may decrease with increasing
$R_{\rm m}$, the field strength at the end of the kinematic regime seems
to be still independent of $R_{\rm m}$.
This is at least qualitatively similar to the behavior of homogeneous
dynamos (BB02, Subramanian 2002).
Furthermore, the peak field strength at the end of the kinematic
phase depends on the strength of the dynamo number, which is also similar
to what is predicted based on homogeneous dynamo theory
(cf.\ Subramanian 2002, Brandenburg \& Subramanian 2005a).

Clearly, further investigations of current helicity fluxes are
warranted to pin down the origin of the apparently unquenched saturation
amplitude of the simulations.
It was already noted by BS04 that the
Vishniac \& Cho flux only accounted for about one quarter of the total
current helicity flux that was determined from the simulations.
This could also indicate
that another perhaps more important component still needs to
be included in the present mean field models.
It is possible that the helicity fluxes discussed by Kleeorin et al.\
(2000, 2002, 2003a,b) may capture the missing components of the
helicity flux, but this remains at this point only speculation.

The present work has shown that there is a threshold of the current
helicity flux above which runaway-type dynamo action  is possible.
It remains a challenge to determine whether in fact all existing
high magnetic Reynolds number dynamos lie in this very same regime.
Clearly, more work is needed to establish whether existing large scale
dynamos without kinetic helicity and just shear (B05) operate in this
supercritical helicity flux regime by the Vishniac \& Cho mechanism,
or whether they work with the shear--current effect, for example.

\acknowledgements
We thank Eric G.\ Blackman for suggestions and comments on the manuscript.
We also thank the organizers of the program ``Magnetohydrodynamics of Stellar
Interiors'' at the Isaac Newton Institute in Cambridge (UK) for creating
a stimulating environment that led to the present work.
The Danish Center for Scientific Computing is acknowledged for granting
time on the Horseshoe cluster.

\begin{appendix}

\section{Vishniac-Cho flux in a shear flow}
\label{app}

We derive here the form of the Vishniac-Cho flux,
introduced in \Eq{helflux1}, assuming that
the underlying turbulence is homogeneous and isotropic
(and weakly or not helical), and that all the anisotropy required
to get a non-zero $\phi_{ijk}$ arises from the influence
of large scale velocity shear. 
We use the two-scale approach of Roberts \& Soward (1975)
whereby one assumes that the correlation tensor of
fluctuating quantities ($\uu$ and $\bb$) vary slowly on
the system scale, say $\RR$.
From Subramanian \& Brandenburg (2004)
we have
\EQ
\phi_{spk}
=-4\tau
\epsilon_{klm}\int k_l k_p v_{ms}(\kk, \RR) \,\dd^3k,
\label{vishflux}
\EN
where
\EQ
v_{ms} =
\int \overline{\hat{u}_m(\kk + \half\KK) \hat{u}_s(-\kk +\half\KK)}.
{\rm e}^{\ii\KK\cdot\RR} \,\dd^3K.
\label{vms}
\EN
Here $\hat{u}_m$ is the Fourier transform of the velocity component
$u_m$. Note that for homogeneous isotropic turbulence $\phi_{spk}$ vanishes.
Now suppose we consider the effect of a weak shear on this turbulence.
Then one can approximate its effect as giving an extra first order
contribution to the velocity, $\uu = \uu^{(0)} + \uu^{(1)}$, where
$\uu^{(0)}$ is the isotropic homogeneous part.
 From the perturbed momentum equation, 
$\uu^{(1)} \approx -\tau_* [ \uu^{(0)}\cdot\nab\meanUU 
+\meanUU\cdot\nab\uu^{(0)} - \nab p]$. Here $p$ is the perturbed pressure
which ensures $\nab\cdot\uu^{(1)}=0$, and $\tau_*$ is some
relaxation time. In Fourier space
one can then write
\EQA
\hat{u}_m^{(1)}(\kk) = &-&\tau_* P_{mj}(\kk)\int  
\big[\ii k'_q\hat{u}_q^{(0)}(\kk -\kk')
\hat{\meanU}_j(\kk') \nonumber \\
&+&\ii(k_q-k'_q)\hat{u}_j^{(0)}(\kk -\kk')\hat{\meanU}_q(\kk') \big] \,\dd^3k',
\label{u1four}
\ENA
where $P_{mj}(\kk) = \delta_{mj} - k_mk_j/k^2$ is the projection operator
which ensures the incompressibility condition on $\uu^{(1)}$. Also 
$\hat{\meanU}_q$ is the Fourier transform of $\meanU_q$.

We now substitute $\uu = \uu^{(0)} + \uu^{(1)}$ in \Eqs{vishflux}{vms},
keeping only terms linear in $\uu^{(1)}$. Let us denote the two terms
on the RHS of \Eq{u1four} as $\hat{u}_s^{(1a)}$ and $\hat{u}_s^{(1b)}$. 
Then on substituting \Eq{u1four} into \Eq{vishflux}, \Eq{vms},
four terms result, schematically of the form, 
Term~I: $\hat{u}_m^{(0)} \hat{u}_s^{(1a)}$,
Term~II: $\hat{u}_m^{(1a)} \hat{u}_s^{(0)}$,
Term~III: $\hat{u}_m^{(0)} \hat{u}_s^{(1b)}$,
Term~IV: $\hat{u}_m^{(1b)} \hat{u}_s^{(0)}$.
The simplification of these terms involve tedious algebra.
We outline the steps for Term~I and simply quote the results
for other terms.
Term~I is given by
\EQA
\phi_{skp}^{\rm I}&=&4\tau\tau_*\epsilon_{klm} \int \big[
k_lk_pP_{mj}(\kk + \KK/2)
\,\ii k'_q
{\rm e}^{i\KK\cdot\RR}\,\hat{\meanU}_j(\kk')
\nonumber \\
&\times& \overline{\hat{u}_q^{(0)}(\kk\!+\!\half\KK\!-\!\kk')
\hat{u}_s^{(0)}(-\kk\!+\!\half\KK)}
\big] \,\dd^3K \,\dd^3k \,\dd^3k'.
\ENA
We change variables to $\KK' = \KK -\kk'$ and integrate over
$\KK'$, keeping in mind that the zeroth order $\uu^{(0)}$ is
homogeneous. Also, since $\meanUU$ only varies slowly
with $\RR$, we retain only up to first derivative in $\meanUU$,
which implies retaining only terms linear in $\kk'$ in the above integrals. 
One then has on evaluating the $\KK'$ and $\kk'$ integrals
\EQA
\phi_{skp}^{\rm I}&=& 4\tau\tau_*\epsilon_{klm}
\nab_q\meanU_m
\int k_lk_p v_{qs}^{(0)} \dd^3k
\nonumber \\
&=& -\frac{4\tau\tau_*{\cal A}}{15}\left[ \epsilon_{ksm} \nab_p\meanU_m 
+ \epsilon_{klm} \nab_l\meanU_m \delta_{ps}\right].
\ENA
Here we have taken the kinetic energy spectrum
for the homogeneous part of the turbulence to be
$v_{qs}^{(0)} = P_{qs}(\kk)E(k)$, 
and done the angular integrals over $\kk$ space.
Also ${\cal A} = \int E(k) k^2 \dd^3k$.
Similarly we get for Term~II:
\EQ
\phi_{skp}^{\rm II}
= -\frac{4\tau\tau_*{\cal A}}{15} \left[\epsilon_{ksm} \nab_m\meanU_p 
+ \epsilon_{klm} \nab_m\meanU_l \delta_{ps}\right]
\EN
and so $\phi_{skp}^{\rm I} + \phi_{skp}^{\rm II} = (8\tau\tau_*{\cal A}/15)\,
\epsilon_{skm}\overline{\sf S}_{mp}$. A similar calculation
can be done for Terms~III and IV to get
$\phi_{skp}^{\rm III} + \phi_{skp}^{\rm IV} = (8\tau\tau_*{\cal A}/3) 
\epsilon_{skm}\overline{\sf S}_{mp}$. Adding all the terms one gets
the expression given in \Eq{helflux2} of the main text,
\EQ
\phi_{ijk}=C_{\rm VC}\,\epsilon_{ijl}\overline{\sf S}_{lk}; \quad 
{\rm with} \quad 
C_{\rm VC} = 16\tau\tau_*{\cal A}/5.
\EN
One can estimate the dimensionless number $C_{\rm VC}$ as
\EQ
C_{\rm VC} = 16\tau \tau_* {\cal A}/5 \sim
{\textstyle{8\over5}}(k_{\rm e} u \tau) (k_{\rm e} u \tau_*)
\sim\mbox{St}^2,
\EN
where we approximated ${\cal A} = \int E(k) k^2\,\dd^3k 
\sim{1\over2}u^2 k_{\rm e}^2$, taken $\tau \sim \tau_*$ and defined a Strouhal number
$\mbox{St} = k_{\rm e} u \tau$.  For a flow dominated by a single scale
$k_{\rm e} \sim k_{\rm f}$, the forcing scale. For a multi scale flow, like
for Kolmogorov turbulence one should also keep the $k$ dependence
of $\tau(k) \propto k^{-2/3}$ say, and $k_{\rm e}$ could be larger by
a logarithmic factor $\ln(k_{\rm d}/k_{\rm f})$ where $k_{\rm d}$ is the dissipative
scale. However in a recent re-formulation
of the dynamical quenching equation using local magnetic helicity
density conservation (Subramanian \& Brandenburg 2005; in preparation)
we have recovered the Vishniac-Cho flux as a magnetic helicity
flux, and in this case $k_{\rm e} \sim k_{\rm f}$. So it is reasonable
to have $\mbox{St}\sim u k_{\rm f}\tau < 1$, and hence $C_{\rm VC} < 8/5$.

\end{appendix}

\vfill\bigskip\noindent\tiny\begin{verbatim}
$Header: /home/brandenb/CVS/tex/mhd/helflux1d/paper.tex,v 1.98 2005/06/24 18:18:44 brandenb Exp $
\end{verbatim}


\begin{thebibliography}{}

\bibitem{}
Arlt, R., Brandenburg, A.\yana{2001}{380}{359}

\bibitem{}
Blackman, E.G., Field, G.B.\yapj{2000}{534}{984}

\bibitem{}
Blackman, E.G., Field, G.B.\ymn{2000}{318}{724}

\bibitem{}
Blackman, E.G., Field, G.B.\ypp{2004}{11}{3264}

\bibitem{}
Blackman, E.G., Brandenburg, A.\yapj{2002}{579}{359} (BB02)

\bibitem{}
Brandenburg, A.\yapj{2001}{550}{824} (B01)

\bibitem{}
Brandenburg, A.\yapj{2005}{625}{539} (B05)

\bibitem{}
Brandenburg, A., Dobler, W.\yana{2001}{369}{329}

\bibitem{}
Brandenburg, A., Sandin, C.\yana{2004}{427}{13} (BS04)

\bibitem{}
Brandenburg, A., Subramanian, K.: 2005a, PhR [arXiv: astro-ph/0405052]

\bibitem{}
Brandenburg, A., Subramanian, K.: 2005b, A\&A [arXiv: astro-ph/0504222]

\bibitem{}
Brandenburg, A., Dobler, W., Subramanian, K.\yan{2002}{323}{99}

\bibitem{}
Brandenburg, A., Haugen, N.E.L., K\"apyl\"a, P.J.,
Sandin, C.\yan{2005}{326}{174} 

\bibitem{}
Field, G.B., Blackman, E.G.\yapj{2002}{572}{685}

\bibitem{}
Gruzinov, A.V., Diamond, P.H.\yprl{1994}{72}{1651}

\bibitem{}
Gruzinov, A.V., Diamond, P.H.\ypp{1995}{2}{1941}

\bibitem{}
Kleeorin, N.I., Ruzmaikin,
A.A.\yjour{1982}{Magne\-to\-hydro\-dynamics}{18}{116}

\bibitem{}
Kleeorin, N., Rogachevskii, I., Ruzmaikin, A.\yana{1995}{297}{159}

\bibitem{}
Kleeorin, N., Moss, D., Rogachevskii, I., Sokoloff, D.\yana{2000}{361}{L5}

\bibitem{}
Kleeorin, N., Moss, D., Rogachevskii, I., Sokoloff, D.\yana{2002}{387}{453}

\bibitem{}
Kleeorin, N., Moss, D., Rogachevskii, I., Sokoloff, D.\yana{2003a}{400}{9}

\bibitem{}
Kleeorin, N., Kuzanyan, K., Moss, D., Rogachevskii, I., Sokoloff, D., 
Zhang, H.\yana{2003b}{409}{1097}

\bibitem{}
Meinel, R., Brandenburg, A.\yana{1990}{238}{369}

\bibitem{}
R\"adler, K.-H.\yjour{1969}{Geod. Geophys. Ver\"off., Reihe II}{13}{131}

\bibitem{}
Roberts, P.H., Soward, A.M.\yan{1975}{296}{49}

\bibitem{}
Rogachevskii, I., Kleeorin, N.\ypre{2003}{68}{036301} 

\bibitem{}
Rogachevskii, I., Kleeorin, N.\ypre{2004}{70}{046310}

\bibitem{}
R\"udiger, G., Kitchatinov, L. L.\spre{2005}

\bibitem{}
Subramanian, K.\yjour{2002}{Bull.\ Astr.\ Soc.\ India}{30}{715}

\bibitem{}
Subramanian, K., Brandenburg, A.\yprl{2004}{93}{205001}

\bibitem{}
Vishniac, E.T., Cho, J.\yapj{2001}{550}{752}

\bibitem{}
Yousef, T.A., Brandenburg, A., R\"udiger, G.\yana{2003}{411}{321}

\end{thebibliography}
\end{document}